# Quantum interface between a transmon qubit and spins of nitrogen-vacancy centers

*Yaowen Hu[†, ‡], Yipu Song[*, †], and Luming Duan[*, †, §]*

[†]Center for Quantum Information, IIIS, Tsinghua University, Beijing 100084, China

[‡]Department of Physics, Tsinghua University, Beijing 100084, China

[§]Department of Physics, University of Michigan, Ann Arbor, Michigan 48109, USA

Hybrid quantum circuits combining advantages of each individual system have provided a promising platform for quantum information processing. Here we propose an experimental scheme to directly couple a transmon qubit to an individual spin in the nitrogen-vacancy (NV) center, with a coupling strength three orders of magnitude larger than that for a single spin coupled to a microwave cavity. This direct coupling between the transmon and the NV center could be utilized to make a transmon bus, leading to a coherently virtual exchange among different single spins. Furthermore, we demonstrate that, by coupling a transmon to a low-density NV ensemble, a SWAP operation between the transmon and NV ensemble is feasible and a quantum non-demolition measurement on the state of NV ensemble can be realized on the cavity-transmon-NV-ensemble hybrid system. Moreover, on this system, a virtual coupling can be achieved between the cavity and NV ensemble, which is much larger in magnitude than the direct coupling between the cavity and the NV ensemble. The photon state in cavity can be thus stored into NV spins more efficiently through this virtual coupling.
Subject Areas: Quantum Physics, Quantum Information

## I. INTRUDUCTION

Quantum information processing has received tremendous attention owing to its potential application in quantum computation and networking [1-3]. Among various kinds of candidates for quantum computing, enormous progress has been made on atomic systems and superconducting qubit systems due to their distinct advantages. Atomic systems, with electron or nuclear spins in the ground-state manifold as the qubits, usually present an excellent coherence time [4,5] since they are well-protected from environmental disturbance. This good isolation, however, is inevitably accompanied with a relatively weak coupling with the outside world [6], which makes it more difficult for coherent manipulation. In contrast, the platform provided by superconducting qubit systems allows for a strong interaction with an external field [7,8], which enables fast control with good scalability [9-12] but leads to a relatively short coherence time.

To make full use of distinctive advantages of these two systems, various studies have been focusing on building a hybrid system to combine the superiority and overcome the drawback of the spin and the superconducting qubits [13-17]. One approach to building a hybrid system is to use a superconducting cavity as a quantum bus. In this design both the spin and the superconducting qubits are coupled to the microwave cavity, and the quantum information can be transferred between the spin and qubit via the quantum bus [13]. To solve the problem of weak interaction between the spin and the microwave cavity, a spin ensemble with N spins is usually used for an increase of coupling strength by a factor of $\sqrt{N}$ [6,13,18]. Due to the low coupling strength between a single spin and cavity, a large number of spins are required to achieve a strong coupling. However, the coherence performance would be degraded because of the spin interaction in high-density ensembles. Besides this cavity-mediated coupling, a direct coupling between a flux qubit and a spin ensemble can be achieved [19,20]. However, the application of the flux qubit in quantum computing is limited due to its short coherence time.

Transmon qubit is the most widely-used qubit in the current superconducting quantum computation architecture due to its relatively long coherence time and low sensitivity to charge noise compared with flux and charge qubits [21]. Unlike the flux qubit, transmon can be strongly coupled to cavity very easily and detected by a non-destructive dispersive readout scheme. NV center has been used as the quantum memory in the hybrid system owing to the attractive properties of extremely long coherence time [5,22]. In this paper, we propose an experimentally feasible hybrid quantum system to directly couple a transmon qubit to spins in NV centers, which has not been exploited before. By directly coupling these two systems, we find that the coupling strength between the transmon and an individual spin is three orders of magnitude larger than that in the cavity-single NV center system, thus greatly reducing the number of NV centers required to achieve strong coupling. The large coupling rate between a single spin and a transmon qubit makes it possible to realize a transmon bus to entangle two or more distant spins, resulting in the transfer of quantum information between spins by the long range virtual exchange. We also investigate a cavity-transmon-NV-ensemble hybrid system and show, by coupling a transmon qubit to a low-density NV ensemble, a SWAP operation between the transmon and the NV ensemble is feasible and a quantum non-demolition measurement on the state of NV ensemble can be realized in this hybrid system.

## II. Quantum interface between a transmon and spins of NV centers
### A. Transmon and spins of NV centers

NV center is an impurity in diamond with the electron spins in S = 1 state. There is a zero magnetic field splitting $\omega_\pm \approx 2.88$ GHz between the state $m_S = 0$ and $m_S = \pm 1$. The spin in NV center is not actually a two level system, but we can induce a splitting between states $m_S = \pm 1$ with an external magnetic field of mT level. Transmon qubit has been historically considered as a special case of Cooper Pair Box (CPB) behaved as an anharmonic oscillator [23]. The simplest architecture of a transmon qubit consists of one Josephson junction (JJ) shunted with a large capacitance. The Hamiltonian of this type of transmon qubit can be written in the phase basis with an offset charge $n_g$ eliminated by a gauge transformation:

$$H_{trans} = -4E_C \frac{\partial^2}{\partial \varphi^2} - E_J \cos\varphi \tag{1}$$

where $E_c$ is the charging energy and $E_J$ is the Josephson energy, $\varphi$ is the phase difference of the wave between the superconductors. This Hamiltonian represents a particle with the position $\varphi$ moving in a cosine potential field. Due to the large capacitance of transmon, it is operated in a regime $E_J \gg E_C$, which leads to a very small fluctuation on phase $\varphi$ for transmon. So we usually deal with this Hamiltonian with the perturbation theory. Expanding the cosine potential in equation (1) gives $H_{trans} = -4E_C \frac{\partial^2}{\partial \varphi^2} - E_J(1 - \frac{\varphi^2}{2} + \frac{\varphi^4}{4!} + \cdots)$. Since in Circuit Quantum Electrodynamics (cQED) the phase difference $\varphi$ and the number of cooper pairs n follow a canonical conjugated commutating relation, we can introduce annihilation and creation operators $b, b^\dagger$ which satisfy the bosonic commutation relation $[b, b^\dagger] = 1$. Expressing $\varphi$ and n as a linear combination: $\varphi = \frac{1}{\sqrt{2}}\left(\frac{8E_C}{E_J}\right)^{\frac{1}{4}}(b + b^\dagger)$, $n = \frac{1}{\sqrt{2}}\left(\frac{E_J}{8E_C}\right)^{\frac{1}{4}}(b - b^\dagger)$, and substituting $\varphi$ and n with b and b† one obtains

$$H_{trans} \approx \hbar\omega_p \left(b^\dagger b + \frac{1}{2}\right) - \frac{E_C}{12}(b + b^\dagger)^4$$

where $\omega_p = \frac{\sqrt{8E_JE_C}}{\hbar}$ is the plasma frequency. Therefore, transmon is like a harmonic oscillator with frequency $\omega_p$ but perturbed by a small nonlinear term $H' = -\frac{E_C}{12}(b + b^\dagger)^4$.

Usually the transmon qubit consists of a superconducting loop with two Josephson junctions (JJ). NV center can be thus placed near the loop. Figure 1 shows a schematic diagram of a cavity-transmon-spin-ensemble hybrid system with a transmon covered by a diamond chip which is located in a superconducting cavity. According to the Josephson relation $I = I_c \sin\theta$, where $I_c$ is the critical current, there will be a current flowing through the transmon. The current will generate a magnetic field, which can be utilized to couple transmon to spins in diamond. The transmon frequency can be tuned in resonance with spin ensemble by an external magnetic field perpendicular to the loop generated by a current bias [24]. Considering a double-Josephson junction (double-JJ) transmon with a transition frequency 3.7 GHz (junction resistance $R_n = 15 \, k\Omega$ and $E_c = 92 \, MHz$), it is straightforward to tune the transition frequency to 2.88 GHz to be resonant with the NV spins with a magnetic field of milli Gauss scale (of the order of 10 mG for a typical geometry of double-JJ transmon), which is much lower than the critical magnetic field for Al thin film [25].

### B. The interaction between a transmon and an individual spin of NV center

We begin with an analysis of the coupling between an individual spin and a transmon with a single JJ. A model of the single-JJ transmon coupling with an individual spin is shown in the inset of Figure 2(a). The Pauli operators are denoted as **τ** for the transmon and **σ** for the NV spins. Here we choose the z axis as the crystalline axis of the NV center and $\sigma_x, \sigma_y, \sigma_z$ are tied to the x-y-z axis of the NV center, while $\tau_x, \tau_y, \tau_z$ have no relationship with the x-y-z axis of the NV center. The spin in the NV center is coupled to the transmon through a magnetic dipole coupling. The interaction term between them is $H_{int} = -\boldsymbol{\mu} \cdot \boldsymbol{B} = \frac{\mu_B g_e}{\hbar} \boldsymbol{S} \cdot \boldsymbol{B}$, where $\mu_B$ is Born magneton, $g_e$ is the g-factor of electron and **B** is the magnetic field generated by the transmon.

The dependence of **B** on the state of transmon can be investigated based on the relation

$\varphi = \frac{1}{\sqrt{2}} \left( \frac{8E_C}{E_J} \right)^{\frac{1}{4}} (b + b^\dagger)$. If the transmon is a perfect harmonic oscillator (i.e., neglecting the nonlinear term), in the qubit space of the transmon, $b, b^\dagger$ would be equivalent to the Pauli lowering and raising operators $\tau_-, \tau_+$. However, with the nonlinear term $H' = -\frac{E_C}{12}(b + b^\dagger)^4$, this equivalence remains a good approximation. The error of the substitution $b \to \tau_-, b^\dagger \to \tau_+$ has been estimated by perturbation theory and shown in detail in Appendix A. We take this substitution and investigate the coupling under this approximation. By substituting $\tau_-, \tau_+$ for $b, b^\dagger$, the current on transmon is

$$I = I_c \sin \varphi \approx I_c \varphi = I_c \frac{1}{\sqrt{2}} \left( \frac{8E_C}{E_J} \right)^{\frac{1}{4}} (\tau_- + \tau_+) = I_c \frac{1}{\sqrt{2}} \left( \frac{8E_C}{E_J} \right)^{\frac{1}{4}} \tau_x$$

Here we use the approximation similar to the analysis of the transmon Hamiltonian that expanding $\sin \varphi$ by ignoring the higher-order terms such as $\frac{\varphi^3}{6}$ as their contribution is very small. For instance, considering a typical parameter of transmon in the regime $E_C/E_J \approx 1/100$, one gets $\varphi = 0.376 \tau_x$, while $\frac{\varphi^3}{6}$ is only $0.009 \tau_x^3$, which is much smaller than the first order. As a result, the magnetic field generated by the current of transmon is proportional to $\tau_x$. For convenience, we denote the field of transmon as $\mathbf{B} = \mathbf{B}_0 \tau_x$ and project $\mathbf{B}, \mathbf{B}_0$ into the x-y-z axis of NV center so that $B_x, B_y, B_z$ and $B_{0_x}, B_{0_y}, B_{0_z}$ are the components along the x,y,z axis, respectively. The interaction can be written as $H_{int} = -\boldsymbol{\mu} \cdot \mathbf{B} = \left( \frac{\sigma_x}{\sqrt{2}} M_x + \frac{\sigma_y}{\sqrt{2}} M_y + \frac{\sigma_z}{2} M_z \right) \tau_x$, where $M_i = \mu_B g_e B_{0_i}$, $(i = x, y, z)$. By putting the Hamiltonian of NV center and transmon together and using a rotating wave approximation, the final Hamiltonian becomes

$$H_{ts}/\hbar = \frac{\omega_t}{2} \tau_z + \frac{\omega_s}{2} \sigma_z + g_{ts} \sigma_+ \tau_- + g_{ts}^* \sigma_- \tau_+,$$

where $g_{ts} = \frac{1}{\sqrt{2}}(M_x - iM_y)/\hbar$ is the coupling strength between the transmon and the spin, and $\omega_t$ and $\omega_s$ are the transition frequency for the transmon and the spin, respectively. The physical picture could be understood as follows: transmon is a nonlinear oscillator, and the current in transmon is like a displacement operator of the oscillator which is proportional to $\tau_x = \tau_- + \tau_+$ under good approximation. So the magnetic field generated by transmon depends on the displacement of this oscillator. The spin interacts with the displacement of the transmon through its generated magnetic field. Alternatively, the coupling can also be understood as the displacement of transmon influenced by the magnetic field generated by the NV spin.

To make the frequency of transmon adjustable, two Josephson junctions forming a SQUID loop are usually used in the transmon design. The inset of Figure 2 (b) shows a schematic diagram of a single spin coupling to a double-JJ transmon. This double-JJ is equivalent to one junction with the flux-dependent Josephson energy [21]:

$$H_J = -(E_{J1} + E_{J2})\left[\cos\frac{\pi\phi}{\phi_0}\cos\varphi + d\sin\frac{\pi\phi}{\phi_0}\sin\varphi\right], \tag{2}$$

where $d = (E_{J2} - E_{J1})/(E_{J1} + E_{J2})$, $\phi$ is the external flux of the loop, and $\varphi = \frac{\theta_1+\theta_2}{2}$ is the phase difference operator of the equivalent one junction. $E_{J1}, \theta_1, E_{J2}, \theta_2$ are the Josephson energy and phase difference for each junction, respectively. Usually, it is preferred to operate with integer flux quanta in the loop, leading to $\sin\frac{\pi\phi}{\phi_0} = 0$, so the second term in equation (2) can be dropped.

Changing the geometry of transmon from one JJ to two JJs would affect the interaction term since there are two separated currents flowing through each junction. Thus, each current of two junctions should be treated separately. Since $\theta_1 - \theta_2 = \frac{2\pi\phi}{\phi_0}$, the currents $I_1, I_2$ on each junction with $\varphi$ expanded to the first order are

$$I_1 = I_{c1}\sin\frac{\pi\phi}{\phi_0} + \varphi I_{c1}\cos\frac{\pi\phi}{\phi_0},$$

$$I_2 = -I_{c2}\sin\frac{\pi\phi}{\phi_0} + \varphi I_{c2}\cos\frac{\pi\phi}{\phi_0}.$$

The term contributing to the coupling is the second one which is proportional to $\cos\frac{\pi\phi}{\phi_0}$. So the coupling would be maximal at an integer flux, which is consistent with the demand on suppressing the second term in equation (2). Operation at the integer flux also has the advantage that the coupling strength would have a minimum fluctuation on this point since it has a zero derivative on flux $\phi$. In the following we assume our double-JJ transmon is operated at the integer flux point, which yields $I_1 = \pm\varphi I_{c1}, I_2 = \pm\varphi I_{c2}$. Based on these two currents, the effective magnetic field and coupling strength $g_{ts}$ can be evaluated. The result indicates that the coupling strength between the transmon and the spin is adjustable by the magnetic flux. Figure 2 (a) and (b) show the estimated coupling strength $g_{ts}$ for an individual spin coupling to a single-JJ transmon and double-JJ transmon, respectively. For the single NV center located at 0.1 μm above the center of the single-JJ transmon with a quantization axis along the transmon orientation (x direction in the Inset of Figure 2 (a) and (b)), the coupling strength is estimated to be $2\pi \times 8$ kHz with a critical current $I_c = 500$ nA. This coupling strength is approximately three orders of magnitude larger than that for a single spin coupling to a microwave cavity. For the case of coupling to a double-JJ transmon, the maximum coupling can be achieved near each junction. This result reveals that even the coupling strength with a single spin can be larger than the decoherence rate of the spin, which is remarkable and important for application of this hybrid system.

### C. The coupling between a transmon and a spin ensemble and quantum non-demolition measurement

The coupling strength of transmon to NV center spin ensemble has a $\sqrt{N}$ enhancement by using an ensemble of N spins that are near resonant to the transmon qubit. The schematic diagram of a transmon coupling to spin ensemble is depicted in Figure 1. The spins in a NV ensemble could have significant inhomogeneous broadening (about 10 MHz scale [26]) and the transmon qubit only couples to those spins that are near resonant with the detuning smaller or comparable with the transmon-spin coupling rate. The total coupling $g_{t-ens}$ can be expressed as $g_{t-ens} = \sqrt{\sum_j |g_j|^2}$, where the summation is over all NV center spins that are near resonant to the transmon qubit. Under a low excitation approximation, the spin ensemble can be treated with a collective spin operator $s = \frac{1}{g_{t-ens}}\sum_j g_j \sigma_-^j$, $s^\dagger = \frac{1}{g_{t-ens}}\sum_j g_j \sigma_+^j$, which satisfy the bosonic creation-annihilation commutation relation $[s, s^\dagger] = 1$. The interaction Hamiltonian between the spin ensemble and the transmon qubit becomes

$$H_{int} = g_{t-ens}(s^\dagger \tau_- + s\tau_+)$$

The coupling strength $g_{t-ens}$ is estimated by summing over all of the inhomogeneous coupling strength $g_{ts}(\boldsymbol{r})$. We assume that the external magnetic field $B_{NV}$ is along the [100] direction of the diamond sample, in line with the direction of transmon (x direction in Figure 1), which has equal components along the four spin axes of the NV center spins in the diamond.

The collective coupling rate $g_{t-ens}$ is shown in Figure 3(a) as a function of the size of the diamond $L_N$ with different densities n of near-resonant NV center spins. Figure 3(b) plots the $g_{t-ens}$ as a function of density n with different dimensions of diamond crystal. These two figures indicate that a coupling strength of 1 MHz can be reached with a crystal size about 4 μm with a low density of NV centers $5 \times 10^{16}\ cm^{-3}$. Considering a typical coherence time ~10 − 100 μs for transmon qubit [14] and ~2 ms for NV center spins [27], strong coupling is readily achievable with a typical crystal size of diamond or NV center density. Comparing with the case of coupling the NV spin ensemble to a microwave cavity mode [26,28] much low spin density is required to reach the strong coupling regime. Note that the inhomogeneous magnetic field from the transmon could also cause an inhomogeneous broadening of the NV spin ensemble, but this broadening is on the order of 10 $kHz$, which is much smaller than the natural broadening of the NV spin ensemble.

Due to the long coherence time of spin ensemble, it is preferred to use spin ensemble as a quantum memory to store the state of transmon in a hybrid quantum circuit. In experiment, a low temperature is required to operate transmon to maintain its superconductivity and fully polarize the NV center spins into the ground state [27]. Thanks to the interaction term $g_{t-ens}(s^\dagger \tau_- + s\tau_+)$, two states $|G\rangle = |g\rangle^N$ and $|B\rangle = s^\dagger |g\rangle^N$ can be used to exchange quantum information with the ground and the excited state of the transmon qubit at zero detuning $\Delta_{t-ens} = 0$. A SWAP gate between the NV spin ensemble and the transmon can be realized by taking a fixed time of interaction $t = \pi/2g_{t-ens}$, which enables us to directly write the quantum state of the transmon qubit into the bright mode of the NV spin ensemble and then retrieve it back to the transmon after a controllable storage time. This coupling between the NV spin ensemble and the transmon qubit also yields an intriguing dispersive readout strategy for the bright mode state of the spin ensemble. In the dispersive regime with $g_{t-ens} \ll \Delta_{t-ens} = \omega_t - \omega_s$, the Hamiltonian of the NV-ensemble-transmon system becomes

$$H_{t-ens}/\hbar \approx \omega_s s^\dagger s + \frac{1}{2}(\omega_t + 2\chi s^\dagger s + \chi)\tau_z \quad (3)$$

where $\chi = g_{t-ens}^2/\Delta_{t-ens}$. In this case, the frequency of transmon qubit depends on the state of spin ensemble, which means that we can read out the state of the NV spin ensemble by detecting the transmon state based on the routine dispersive readout via the cavity. Figure 4 (a) shows the measurement strategy and an energy level diagram of the NV-ensemble-transmon system is illustrated in (b). The transition frequency of transmon depends on the state of NV ensemble due to the dispersive interaction between the transmon and the NV spin ensemble. The frequency shifts by $2\chi$ when the NV ensemble is excited from the ground state $|G\rangle$ to the excited bright mode state $s^\dagger|G\rangle$. For the measurement process, the transmon and the spin ensemble are initially prepared in a ground stat $|\uparrow G\rangle$. Next, the transmon is excited by a pump pulse at a frequency of $\omega_t + \chi$. The transmon frequency is then tuned in resonance with the spin ensemble with an external flux generated by a current bias, resulting in a SWAP gate to transfer quantum state to the spin ensemble. The transmon frequency is tuned back afterwards to turn off the exchange coupling between the transmon and the spin ensemble. The final state of spin ensemble is determined by a $\pi$ pulse or $\frac{\pi}{2}$ ramsey pulse sequence on the transmon at the frequency of $\omega_t + \chi$, followed by a readout pulse on the cavity to probe the state of the transmon. If the SWAP gate is successfully accomplished, the state of spin ensemble will be changed from $|G\rangle$ to $s^\dagger|G\rangle$. This leads to a frequency shift of $2\chi$ for the transmon, resulting in probing a ground state of transmon when the pump pulse is applied on transmon at the frequency of $\omega_t + \chi$. The higher-order excited states of the bright mode, such as $(s^\dagger)^2|G\rangle$, can be similarly detected by probing the transmon state with the pump pulse at a frequency of $\omega_t + 5\chi$. This measurement needs to be operated in the dispersive regime to ensure the validity of the dispersive Hamiltonian, and it is a quantum-non-demolition measurement since $[s^\dagger s \tau_z, H_{t-ens}] = 0$ [29]. Note that the detuning $\Delta_{ts}$ does not break the rotating wave approximation (RWA).

To show the feasibility of this manipulation and detection scheme, let us take some typical experimental parameters. The frequencies of the transmon, the spin ensemble and the cavity are taken to be 3.27 GHz, 2.88 GHz and 5 GHz, respectively. Assuming the double-JJ transmon has a typical relaxation time $T_1 \sim 20\ \mu s$ with a coupling strength $g_{t-ens} = 15\ \text{MHz}$ to the spin ensemble and $g_{t-c} = 80\ \text{MHz}$ to the cavity, the frequency shift of the transmon is estimated to be $2\chi \approx 1.15\ \text{MHz}$ when the state of spin ensemble changes from $|G\rangle$ to $s^\dagger|G\rangle$, which is large enough to distinguish the state of spin ensemble by dispersively probing the state of transmon with a pump microwave pulse at the frequency of $\omega_t + \chi$.

The state stored in the bright mode would leak into the dark mode due to the inhomogeneous broadening of the NV spin ensemble. To accomplish this QND measurement, the microwave pulse must be fast enough to finish measurement before the state leaks into the dark mode, which is feasible since the leakage into dark modes takes place over a period of time, on the order of free induction decay time of a few μs [30-33]. If we are starting a measurement with the state already stored in the dark mode, refocusing techniques can be utilized to actively restore the state into the bright mode. Then QND measurement can be accomplished as described above. This leakage to dark mode also benefits the storage of state since the dark states are unaffected by spontaneous emission caused by the coupling of spins to the transmon. In using of the dark modes, our hybrid circuit is similar to and compatible with another protocol of writing and reading states with NV spin ensemble as quantum memory described in [27,32,34], where cavity is used as a quantum bus between the NV ensemble and the transmon. Here, our protocol accomplishes state exchange between the processor (transmon) and the memory (NV spin ensemble) through direct coupling between them with a much larger coupling rate without using of a cavity as the inter-media.

### D. Virtual exchange and transmon bus

As discussed before, the coupling strength between a single-JJ transmon and an individual single spin is about 8 kHz, which is not strong enough to coherently transfer quantum information between transmon and spin. However, this coupling can yield a coherent information transfer between two spins via a virtual exchange with the transmon as an intermediary bus. Figure 5 (a) shows the schematic diagram of a hybrid system composed of two spins coupling to transmon. The Hamiltonian of this system is:

$$H_{s-t-s}/\hbar = \frac{\omega_t}{2}\tau_z + \frac{\omega_{s1}}{2}\sigma_{z1} + \frac{\omega_{s2}}{2}\sigma_{z2} + g_{ts1}(\sigma_1^+\tau_- + \sigma_1^-\tau_+) + g_{ts2}(\sigma_2^+\tau_- + \sigma_2^-\tau_+) \quad (4)$$

In the dispersive regime, $g_{ts1} \ll \Delta_{ts1} = \omega_t - \omega_{s1}$, $g_{ts2} \ll \Delta_{ts2} = \omega_t - \omega_{s2}$, we can apply an unitary transformation $U = Exp[-\frac{g_{ts1}}{\Delta_{ts1}}(\sigma_1^+\tau_- - \sigma_1^-\tau_+) - \frac{g_{ts2}}{\Delta_{ts2}}(\sigma_2^+\tau_- - \sigma_2^-\tau_+)]$ to the Hamiltonian $H_{s-t-s}$ and obtain a new Hamiltonian:

$$H_{s-t-s}/\hbar \approx \frac{1}{2}\left(\omega_t + \frac{g_{ts1}^2}{\Delta_{ts1}} + \frac{g_{ts2}^2}{\Delta_{ts2}}\right)\tau_z + \frac{1}{2}\left(\omega_{s1} - \frac{g_{ts1}^2}{\Delta_{ts1}}\right)\sigma_{z1} + \frac{1}{2}\left(\omega_{s2} - \frac{g_{ts2}^2}{\Delta_{ts2}}\right)\sigma_{z2}$$

$$+J(\sigma_1^+ \sigma_2^- + \sigma_1^- \sigma_2^+)\tau_z \quad (5)$$

where $J = \frac{g_{ts1}g_{ts2}}{2}(\frac{1}{\Delta_{ts1}} + \frac{1}{\Delta_{ts2}})$. In the dispersive regime, apart from the frequency shifts to the transmon and the spins, a new interaction term $J(\sigma_1^+ \sigma_2^- + \sigma_1^- \sigma_2^+)\tau_z$ emerges, representing a virtual exchange between the two spins induced by the coupling to the common transmon bus. When two spins are tuned into resonance, a SWAP gate between them can be achieved via this virtual exchange at the interaction time $t = \pi/2J$. Via this virtual exchange, spins are protected from the transmon-induced loss by a reduction factor of $\frac{g_{ts1,2}^2}{\Delta_{ts1,2}^2}$. The coupling mediated by the transmon is of significantly longer range compared with the direct dipole interaction between the two spins. This visual coupling can also be used to aid interaction between many spins near the transmon.

The virtual exchange can also be applied to a cavity-transmon-spin-ensemble system:

$$H_{c-t-ens}/\hbar = \omega_r a^\dagger a + \frac{\omega_t}{2}\tau_z + \omega_s s^\dagger s + g_{tc}(a^\dagger \tau_- + a\tau_+) + g_{t-ens}(s^\dagger \tau_- + s\tau_+) \quad (6)$$

In a dispersive regime $g_{tc} \ll \Delta_{tc} = \omega_t - \omega_r, g_{t-ens} \ll \Delta_{t-ens} = \omega_t - \omega_s$, the Hamiltonian can be simplified by an unitary transformation $U = Exp[-\frac{g_{tc}}{\Delta_{tc}}(a^\dagger \tau_- - a\tau_+) - \frac{g_{t-ens}}{\Delta_{t-ens}}(s^\dagger \tau_- - s\tau_+)]$ to the form:

$$H_{c-t-ens}/\hbar = \frac{1}{2}\left(\omega_t + \frac{2g_{tc}^2}{\Delta_{tc}}a^\dagger a + \frac{g_{tc}^2}{\Delta_{tc}} + \frac{2g_{t-ens}^2}{\Delta_{t-ens}}s^\dagger s + \frac{g_{t-ens}^2}{\Delta_{t-ens}}\right)\tau_z + \omega_s s^\dagger s + \omega_r a^\dagger a$$

$$+g_{virtual}(a^\dagger s + as^\dagger)\tau_z \quad (7)$$

where $g_{virtual} = \frac{g_{tc}g_{t-ens}}{2}(\frac{1}{\Delta_{tc}} + \frac{1}{\Delta_{t-ens}})$. Similar to the case of spin-transmon-spin system, the Hamiltonian shows there is a virtual interaction between the cavity and spin ensemble via the transmon bus. Figure 5 (c) illustrates the energy level of this cavity-transmon-spin-ensemble hybrid system. The state exchange is feasible between the yellow and the blue energy levels. If the system is prepared in the state $|1 \downarrow G\rangle$ at t=0, when the virtual exchange is turned on, the state of system experiences an evolution as $|\psi(t)\rangle = \cos(g_{virtual}t)|1 \downarrow G\rangle + \sin(g_{virtual}t)|0 \downarrow B\rangle$, achieving a complete state transfer at $t = \pi/2g_{virtual}$. This virtual coupling is much larger than the direct coupling of cavity to spin ensemble under the same number of effective spins. For instance, considering $g_{tc} = g_{t-ens} = 10MHz$ and detuning $g_{tc}/\Delta_{tc} = g_{t-ens}/\Delta_{t-ens} = 1/10$, we estimate the virtual coupling $g_{virtual} = 1 MHz = \frac{1}{10} g_{t-ens}$. While, with the same number of spins, the directly coupling rate $g_{c-ens}$ between the cavity and the spin ensemble is only $g_{c-ens} = \frac{1}{1000} g_{t-ens}$. The reason is that, the coupling strength $g_{ts}$ is three orders of magnitude larger than $g_{cs}$ for a single spin coupling to a microwave cavity. Thus, much less number of spins are required to achieve the strong coupling regime for coupling NV ensemble to cavity via the transmon instead of directly coupling NV ensemble to cavity.

## III. CONCLUSIONS

In summary, we have proposed a new hybrid system of directly coupling transmon qubit to a NV center/spin ensemble of NV centers. We estimate the coupling strength between the transmon qubit to a NV center/NV center spin ensemble under different coupling configurations. The coupling rate between the transmon and NV spin is three orders of magnitude larger than that for a single spin coupling to a microwave cavity, which can be used to make a transmon bus, leading to coherent virtual exchange interaction among different single spins. We also demonstrate that, by using a low-density NV spin ensemble, a SWAP operation between the transmon and the NV spin ensemble is feasible and a quantum non-demolition measurement on the state of NV ensemble can be realized on the transmon-NV-ensemble hybrid system. Finally, we investigate the cavity-transmon-NV-ensemble system, and show that coherent information transfer can be achieved between cavity and NV spin ensemble by virtual exchange mediated through the transmon, which is much stronger than the direct coupling between the cavity and the NV spin ensemble. Our proposal of coupling transmon qubit to the NV center spin is feasible with the experimental technology. The parameter estimation is based on typical experimental values. The proposed idea here can also be extended to other spin systems, including, for instance, spins of molecular nanomagnets and phosphorus atoms in silicon, with the potential advantage of combining the long coherence time of spin systems with fast and convenient quantum information processing offered by the transmon qubits.

**ACKNOWLEDGMENTS**

We thank Yukai Wu and Hongyi Zhang for helpful discussions. This work was supported by the State's Key Project of Research and Development Plan under Grant No. 2016YFA0301902.

**Appendix A: Error estimation of substitution** $b \to \tau_-, b^\dagger \to \tau_+$

To estimate the error of substitution $b \to \tau_-, b^\dagger \to \tau_+$, the state of transmon is calculated by perturbation theory. As shown in the main text, the Hamiltonian of transmon is:

$$H_{trans} \approx \hbar\omega_p \left(b^\dagger b + \frac{1}{2}\right) + H'$$

$$H' = -\frac{E_C}{12}(b+b^\dagger)^4 = -\frac{E_C}{12}(b^4 + b^{\dagger 4} + b^2(4\tilde{n}-2) + b^{\dagger 2}(4\tilde{n}+6) + 6\tilde{n}^2 + 6\tilde{n} + 3)$$

where $\tilde{n} = b^\dagger b$. In the following, we denote the eigenstate of $b^\dagger b$ as $|n\rangle$ and take the ground and excited state of transmon as $|\downarrow\rangle, |\uparrow\rangle$. By applying the time-independent perturbation theory, we obtain

$$|0\rangle + |0\rangle^{(1)} = |0\rangle - \frac{E_C}{12}\frac{\langle 2|b^{\dagger 2}(4\tilde{n}+6)|0\rangle}{2\hbar\omega_p}|2\rangle - \frac{E_C}{12}\frac{\langle 4|b^{\dagger 4}|0\rangle}{4\hbar\omega_p}|4\rangle$$

$$|1\rangle + |1\rangle^{(1)} = |1\rangle - \frac{E_C}{12}\frac{\langle 3|b^{\dagger 2}(4\tilde{n}+6)|1\rangle}{2\hbar\omega_p}|3\rangle - \frac{E_C}{12}\frac{\langle 5|b^{\dagger 4}|1\rangle}{4\hbar\omega_p}|5\rangle$$

After some simplifications, we have

$$|0\rangle + |0\rangle^{(1)} = |0\rangle - \frac{6\sqrt{2}}{24}\sqrt{\frac{E_C}{8E_J}}|2\rangle - \frac{\sqrt{24}}{48}\sqrt{\frac{E_C}{8E_J}}|4\rangle$$

$$|1\rangle + |1\rangle^{(1)} = |1\rangle - \frac{10\sqrt{6}}{24}\sqrt{\frac{E_C}{8E_J}}|3\rangle - \frac{\sqrt{120}}{48}\sqrt{\frac{E_C}{8E_J}}|5\rangle$$

By taking $\frac{E_J}{E_C} = 100$ and renormalizing the perturbed states, we get

$$|\downarrow\rangle = \frac{|0\rangle + |0\rangle^{(1)}}{||0\rangle + |0\rangle^{(1)}|} = 0.9999|0\rangle - 0.0125|2\rangle - 0.0036|4\rangle$$

$$|\uparrow\rangle = \frac{|1\rangle + |1\rangle^{(1)}}{||1\rangle + |1\rangle^{(1)}|} = 0.9993|1\rangle - 0.0361|3\rangle - 0.0080|5\rangle$$

So the matrix element of $b + b^\dagger$ can be calculated on the perturbed basis $|\downarrow\rangle, |\uparrow\rangle$

$$b + b^\dagger = \begin{pmatrix} 0 & 0.983 \\ 0.983 & 0 \end{pmatrix} \approx \begin{pmatrix} 0 & 1 \\ 1 & 0 \end{pmatrix} = \tau_- + \tau_+$$

The matrix element error between $(b + b^\dagger)_{ij}$ and $(\tau_- + \tau_+)_{ij}$ is

$$\left| \frac{(b + b^\dagger)_{12} - (\tau_- + \tau_+)_{12}}{(\tau_- + \tau_+)_{12}} \right| = 1.7\%$$

$$\left| \frac{(b + b^\dagger)_{21} - (\tau_- + \tau_+)_{21}}{(\tau_- + \tau_+)_{21}} \right| = 1.7\%$$

Accordingly, this error is small enough to treat $b + b^\dagger$ as $\tau_x$ approximately and it can be further suppressed with larger $E_J/E_c$.

**Appendix B: Derivation of the virtual exchange**

Here we illustrate the detailed derivation of the virtual exchange interaction term in equations (5) and (7) in the main text.

For the case of two spins coupling to a transmon, the original Hamiltonian of the system (equation (4)) is:

$$H_{s-t-s}/\hbar = \frac{\omega_t}{2}\tau_z + \frac{\omega_{s1}}{2}\sigma_{z1} + \frac{\omega_{s2}}{2}\sigma_{z2} + g_{ts1}(\sigma_1^+\tau_- + \sigma_1^-\tau_+) + g_{ts2}(\sigma_2^+\tau_- + \sigma_2^-\tau_+)$$

Applying an unitary transformation $U = Exp[-\frac{g_{ts1}}{\Delta_{ts1}}(\sigma_1^+\tau_- - \sigma_1^-\tau_+) - \frac{g_{ts2}}{\Delta_{ts2}}(\sigma_2^+\tau_- - \sigma_2^-\tau_+)]$ to the Hamiltonian with $\Delta_{ts1} = \omega_t - \omega_{s1}, \Delta_{ts2} = \omega_t - \omega_{s2}$, we can get a new Hamiltonian $UHU^\dagger$. Using the Hausdorff expansion to the second order with $\frac{g_{ts1}}{\Delta_{ts1}}, \frac{g_{ts2}}{\Delta_{ts2}}$ as small parameters,

$$e^{-C}He^{C} = H + [H, C] + \frac{1}{2}[[H, C], C] + \cdots,$$

and denoting $X_1 = \sigma_1^+\tau_- - \sigma_1^-\tau_+$ and $X_2 = \sigma_2^+\tau_- - \sigma_2^-\tau_+$, we get the transformed Hamiltonian:

$$UHU^\dagger = e^{-\left(\frac{g_{ts1}}{\Delta_{ts1}}X_1 + \frac{g_{ts2}}{\Delta_{ts2}}X_2\right)} H e^{\left(\frac{g_{ts1}}{\Delta_{ts1}}X_1 + \frac{g_{ts2}}{\Delta_{ts2}}X_2\right)}$$

$$\approx H + \frac{g_{ts1}}{\Delta_{ts1}}[H, X_1] + \frac{g_{ts2}}{\Delta_{ts2}}[H, X_2]$$

$$+ \frac{1}{2}\frac{g_{ts1}}{\Delta_{ts1}}\left[\left[H, \frac{g_{ts1}}{\Delta_{ts1}}X_1\right], X_1\right] + \frac{1}{2}\frac{g_{ts1}}{\Delta_{ts1}}\left[\left[H, \frac{g_{ts2}}{\Delta_{ts2}}X_2\right], X_1\right]$$

$$+\frac{1}{2}\frac{g_{ts2}}{\Delta_{ts2}}\left[\left[H,\frac{g_{ts1}}{\Delta_{ts1}}X_1\right],X_2\right]+\frac{1}{2}\frac{g_{ts2}}{\Delta_{ts2}}\left[\left[H,\frac{g_{ts2}}{\Delta_{ts2}}X_2\right],X_2\right]$$

The commutating relation used for this derivation is:

$$[\tau_z, X_1] = -2\sigma_1^+\tau_- - 2\sigma_1^-\tau_+ \qquad [\tau_z, X_2] = -2\sigma_2^+\tau_- - 2\sigma_2^-\tau_+$$
$$[\sigma_{z1}, X_1] = 2\sigma_1^+\tau_- + 2\sigma_1^-\tau_+ \qquad [\sigma_{z1}, X_2] = 0$$
$$[\sigma_{z2}, X_1] = 0 \qquad [\sigma_{z2}, X_2] = 2\sigma_2^+\tau_- + 2\sigma_2^-\tau_+$$
$$[\sigma_1^+\tau_- + \sigma_1^-\tau_+, X_1] = \tau_z - \sigma_{z1} \qquad [\sigma_1^+\tau_- + \sigma_1^-\tau_+, X_2] = (\sigma_1^+\sigma_2^- + \sigma_1^-\sigma_2^+)\tau_z$$
$$[\sigma_2^+\tau_- + \sigma_2^-\tau_+, X_1] = (\sigma_1^+\sigma_2^- + \sigma_1^-\sigma_2^+)\tau_z \qquad [\sigma_2^+\tau_- + \sigma_2^-\tau_+, X_2] = \tau_z - \sigma_{z2}$$

We thus obtain:

$$\frac{g_{ts1}}{\Delta_{ts1}}[H_{s-t-s}, X_1] + \frac{g_{ts2}}{\Delta_{ts2}}[H_{s-t-s}, X_2]$$

$$= -g_{ts1}(\sigma_1^+\tau_- + \sigma_1^-\tau_+) - g_{ts2}(\sigma_2^+\tau_- + \sigma_2^-\tau_+) + \frac{g_{ts1}^2}{\Delta_{ts1}}(\tau_z - \sigma_{z1})$$

$$+ \frac{g_{ts2}^2}{\Delta_{ts2}}(\tau_z - \sigma_{z2}) + g_{ts1}g_{ts2}\left(\frac{1}{\Delta_{ts1}} + \frac{1}{\Delta_{ts2}}\right)(\sigma_1^+\sigma_2^- + \sigma_1^-\sigma_2^+)\tau_z$$

And

$$\frac{1}{2}\frac{g_{ts1}}{\Delta_{ts1}}\left[\left[H_{s-t-s},\frac{g_{ts1}}{\Delta_{ts1}}X_1\right],X_1\right] + \frac{1}{2}\frac{g_{ts1}}{\Delta_{ts1}}\left[\left[H_{s-t-s},\frac{g_{ts2}}{\Delta_{ts2}}X_2\right],X_1\right]$$

$$+\frac{1}{2}\frac{g_{ts2}}{\Delta_{ts2}}\left[\left[H_{s-t-s},\frac{g_{ts1}}{\Delta_{ts1}}X_1\right],X_2\right] + \frac{1}{2}\frac{g_{ts2}}{\Delta_{ts2}}\left[\left[H_{s-t-s},\frac{g_{ts2}}{\Delta_{ts2}}X_2\right],X_2\right]$$

$$\approx -\frac{1}{2}\frac{g_{ts1}^2}{\Delta_{ts1}}(\tau_z - \sigma_{z1}) - \frac{g_{ts1}g_{ts2}}{2\Delta_{ts1}}(\sigma_1^+\sigma_2^- + \sigma_1^-\sigma_2^+)\tau_z$$

$$-\frac{g_{ts1}g_{ts2}}{2\Delta_{ts2}}(\sigma_1^+\sigma_2^- + \sigma_1^-\sigma_2^+)\tau_z - \frac{1}{2}\frac{g_{ts2}^2}{\Delta_{ts2}}(\tau_z - \sigma_{z2})$$

It is noted here that we only keep the term $\frac{g}{\Delta}$ and ignore the higher order term like $\frac{g^2}{\Delta^2}$.

By combining the above results we can get the transformed Hamiltonian in equation (5) in the main text:

$$H_{s-t-s}/\hbar \approx \frac{1}{2}\left(\omega_t + \frac{g_{ts1}^2}{\Delta_{ts1}} + \frac{g_{ts2}^2}{\Delta_{ts2}}\right)\tau_z + \frac{1}{2}\left(\omega_{s1} - \frac{g_{ts1}^2}{\Delta_{ts1}}\right)\sigma_{z1} + \frac{1}{2}\left(\omega_{s2} - \frac{g_{ts2}^2}{\Delta_{ts2}}\right)\sigma_{z2}$$
$$+ J(\sigma_1^+\sigma_2^- + \sigma_1^-\sigma_2^+)\tau_z$$

where $J = \frac{g_{ts1}g_{ts2}}{2}\left(\frac{1}{\Delta_{ts1}} + \frac{1}{\Delta_{ts2}}\right)$.

We use the similar procedure as shown above to derive the transformed Hamiltonian for the system of cavity-transmon-NV-ensemble. The Hamiltonian of this hybrid system is:

$$H_{c-t-ens}/\hbar = \omega_r a^\dagger a + \frac{\omega_t}{2}\tau_z + \omega_s s^\dagger s + g_{tc}(a^\dagger \tau_- + a\tau_+) + g_{t-ens}(s^\dagger \tau_- + s\tau_+)$$

We use the transformation $U = Exp[-\frac{g_{tc}}{\Delta_{tc}}(a^\dagger \tau_- - a\tau_+) - \frac{g_{t-ens}}{\Delta_{t-ens}}(s^\dagger \tau_- - s\tau_+)]$, where $\Delta_{tc} = \omega_t - \omega_r$ and $\Delta_{t-ens} = \omega_t - \omega_s$. By denoting $Y_1 = a^\dagger \tau_- - a\tau_+$ and $Y_2 = s^\dagger \tau_- - s\tau_+$, we get the transformed Hamiltonian,

$$UHU^\dagger = e^{-\left(\frac{g_{tc}}{\Delta_{tc}}Y_1 + \frac{g_{t-ens}}{\Delta_{t-ens}}Y_2\right)} H e^{\left(\frac{g_{tc}}{\Delta_{tc}}Y_1 + \frac{g_{t-ens}}{\Delta_{t-ens}}Y_2\right)}$$

$$\approx H + \frac{g_{tc}}{\Delta_{tc}}[H, Y_1] + \frac{g_{t-ens}}{\Delta_{t-ens}}[H, Y_2]$$

$$+ \frac{1}{2}\frac{g_{tc}}{\Delta_{tc}}\left[\left[H, \frac{g_{tc}}{\Delta_{tc}}Y_1\right], Y_1\right] + \frac{1}{2}\frac{g_{tc}}{\Delta_{tc}}\left[\left[H, \frac{g_{t-ens}}{\Delta_{t-ens}}Y_2\right], Y_1\right]$$

$$+ \frac{1}{2}\frac{g_{t-ens}}{\Delta_{t-ens}}\left[\left[H, \frac{g_{tc}}{\Delta_{tc}}Y_1\right], Y_2\right] + \frac{1}{2}\frac{g_{t-ens}}{\Delta_{t-ens}}\left[\left[H, \frac{g_{t-ens}}{\Delta_{t-ens}}Y_2\right], Y_2\right]$$

Using the following commutating relations:

$[\tau_z, Y_1] = -2a^\dagger \tau_- - 2a\tau_+$ $\qquad$ $[\tau_z, Y_2] = -2s^\dagger \tau_- - 2s\tau_+$

$[a^\dagger a, Y_1] = a^\dagger \tau_- + a\tau_+$ $\qquad$ $[a^\dagger a, Y_2] = 0$

$[s^\dagger s, Y_1] = 0$ $\qquad$ $[s^\dagger s, Y_2] = s^\dagger \tau_- + s\tau_+$

$[a^\dagger \tau_- + a\tau_+, Y_1] = 2a^\dagger a\tau_z + \tau_z + 1$ $\qquad$ $[a^\dagger \tau_- + a\tau_+, Y_2] = (a^\dagger s + as^\dagger)\tau_z$

$[s^\dagger \tau_- + s\tau_+, Y_1] = (a^\dagger s + as^\dagger)\tau_z$ $\qquad$ $[s^\dagger \tau_- + s\tau_+, Y_2] = 2s^\dagger s\tau_z + \tau_z + 1$

We get results:

$$\frac{g_{tc}}{\Delta_{tc}}[H_{c-t-ens}, Y_1] + \frac{g_{t-ens}}{\Delta_{t-ens}}[H_{c-t-ens}, Y_2] = -g_{tc}(a^\dagger \tau_- + a\tau_+) - g_{t-ens}(s^\dagger \tau_- + s\tau_+)$$

$$+ g_{tc}g_{t-ens}\left(\frac{1}{\Delta_{tc}} + \frac{1}{\Delta_{t-ens}}\right)(a^\dagger s + as^\dagger)\tau_z + \frac{g_{tc}^2}{\Delta_{tc}}(2a^\dagger a\tau_z + \tau_z + 1) + \frac{g_{t-ens}^2}{\Delta_{t-ens}}(2s^\dagger s\tau_z + \tau_z + 1)$$

And

$$+ \frac{1}{2}\frac{g_{tc}}{\Delta_{tc}}\left[\left[H_{c-t-ens}, \frac{g_{tc}}{\Delta_{tc}}Y_1\right], Y_1\right] + \frac{1}{2}\frac{g_{tc}}{\Delta_{tc}}\left[\left[H_{c-t-ens}, \frac{g_{t-ens}}{\Delta_{t-ens}}Y_2\right], Y_1\right]$$

$$+ \frac{1}{2}\frac{g_{t-ens}}{\Delta_{t-ens}}\left[\left[H_{c-t-ens}, \frac{g_{tc}}{\Delta_{tc}}Y_1\right], Y_2\right] + \frac{1}{2}\frac{g_{t-ens}}{\Delta_{t-ens}}\left[\left[H_{c-t-ens}, \frac{g_{t-ens}}{\Delta_{t-ens}}Y_2\right], Y_2\right]$$

$$= -\frac{g_{tc}^2}{2\Delta_{tc}}(2a^\dagger a \tau_z + \tau_z + 1) - \frac{g_{tc}g_{t-ens}}{2\Delta_{tc}}(a^\dagger s + as^\dagger)\tau_z$$

$$-\frac{g_{tc}g_{t-ens}}{2\Delta_{t-ens}}(a^\dagger s + as^\dagger)\tau_z - \frac{g_{t-ens}^2}{2\Delta_{t-ens}}(2s^\dagger s \tau_z + \tau_z + 1)$$

By keeping the term $\frac{g}{\Delta}$ and ignoring the higher order term like $\frac{g^2}{\Delta^2}$, we finally get the transformed Hamiltonian for the cavity-transmon-NV-ensemble system,

$$H_{c-t-ens}/\hbar = \frac{1}{2}\left(\omega_t + \frac{2g_{tc}^2}{\Delta_{tc}}a^\dagger a + \frac{g_{tc}^2}{\Delta_{tc}} + \frac{2g_{t-ens}^2}{\Delta_{t-ens}}s^\dagger s + \frac{g_{t-ens}^2}{\Delta_{t-ens}}\right)\tau_z + \omega_s s^\dagger s + \omega_r a^\dagger a$$

$$+ g_{virtual}(a^\dagger s + as^\dagger)\tau_z$$

where $g_{virtual} = \frac{g_{tc}g_{t-ens}}{2}\left(\frac{1}{\Delta_{tc}} + \frac{1}{\Delta_{t-ens}}\right)$.

**References:**

*Corresponding Author

E-mail: lmduan@umich.edu, ypsong@mail.tsinghua.edu.cn

[1]     N. Gisin, G. Ribordy, W. Tittel, and H. Zbinden, Rev. Mod. Phys. **74**, 145 (2002).

[2]     M. H. Devoret and R. J. Schoelkopf, Science **339**, 1169 (2013).

[3]     J. Clarke and F. K. Wilhelm, Nature **453**, 1031 (2008).

[4]     C. F. Roos, G. P. T. Lancaster, M. Riebe, H. Häffner, W. Hänsel, S. Gulde, C. Becher, J. Eschner, F. Schmidt-Kaler, and R. Blatt, Phys. Rev. Lett. **92**, 220402 (2004).

[5]     G. Balasubramanian, P. Neumann, D. Twitchen, M. Markham, R. Kolesov, N. Mizuochi, J. Isoya, J. Achard, J. Beck, J. Tissler, V. Jacques, P. R. Hemmer, F. Jelezko, and J. Wrachtrup, Nature Mater. **8**, 383 (2009).

[6]     M. D. Lukin, Rev. Mod. Phys. **75**, 457 (2003).

[7]     J. Q. You and F. Nori, Phys. Rev. B **68**, 064509 (2003).

[8]     A. Wallraff, D. I. Schuster, A. Blais, L. Frunzio, R. S. Huang, J. Majer, S. Kumar, S. M. Girvin, and R. J. Schoelkopf, Nature **431**, 162 (2004).

[9]     J. Majer, J. M. Chow, J. M. Gambetta, J. Koch, B. R. Johnson, J. A. Schreier, L. Frunzio, D. I. Schuster, A. A. Houck, A. Wallraff, A. Blais, M. H. Devoret, S. M. Girvin, and R. J. Schoelkopf, Nature **449**, 443 (2007).

[10]    R. Barends, J. Kelly, A. Megrant, D. Sank, E. Jeffrey, Y. Chen, Y. Yin, B. Chiaro, J. Mutus, C. Neill, P. O'Malley, P. Roushan, J. Wenner, T. C. White, A. N. Cleland, and J. M. Martinis, Phys. Rev. Lett. **111**, 080502 (2013).

[11]    E. Jeffrey, D. Sank, J. Y. Mutus, T. C. White, J. Kelly, R. Barends, Y. Chen, Z. Chen, B. Chiaro, A.

Dunsworth, A. Megrant, P. J. J. O'Malley, C. Neill, P. Roushan, A. Vainsencher, J. Wenner, A. N. Cleland, and J. M. Martinis, Phys. Rev. Lett. **112**, 190504 (2014).

[12]    J. Kelly, R. Barends, A. G. Fowler, A. Megrant, E. Jeffrey, T. C. White, D. Sank, J. Y. Mutus, B. Campbell, Y. Chen, Z. Chen, B. Chiaro, A. Dunsworth, E. Lucero, M. Neeley, C. Neill, P. J. J. O'Malley, C. Quintana, P. Roushan, A. Vainsencher, J. Wenner, and J. M. Martinis, Phys. Rev. A **94**, 032321 (2016).

[13]    Y. Kubo, C. Grezes, A. Dewes, T. Umeda, J. Isoya, H. Sumiya, N. Morishita, H. Abe, S. Onoda, T. Ohshima, V. Jacques, A. Dreau, J. F. Roch, I. Diniz, A. Auffeves, D. Vion, D. Esteve, and P. Bertet, Phys. Rev. Lett. **107**, 220501 (2011).

[14]    Z.-L. Xiang, S. Ashhab, J. Q. You, and F. Nori, Rev. Mod. Phys. **85**, 623 (2013).

[15]    A. Imamoglu, Phys. Rev. Lett. **102**, 083602 (2009).

[16]    X. Zhu, Y. Matsuzaki, R. Amsuss, K. Kakuyanagi, T. Shimo-Oka, N. Mizuochi, K. Nemoto, K. Semba, W. J. Munro, and S. Saito, Nat. Commun. **5**, 3424 (2014).

[17]    D. I. Schuster, A. P. Sears, E. Ginossar, L. DiCarlo, L. Frunzio, J. J. L. Morton, H. Wu, G. A. D. Briggs, B. B. Buckley, D. D. Awschalom, and R. J. Schoelkopf, Phys. Rev. Lett. **105**, 140501 (2010).

[18]    R. Amsüss, C. Koller, T. Nöbauer, S. Putz, S. Rotter, K. Sandner, S. Schneider, M. Schramböck, G. Steinhauser, H. Ritsch, J. Schmiedmayer, and J. Majer, Phys. Rev. Lett. **107**, 060502 (2011).

[19]    X. Zhu, S. Saito, A. Kemp, K. Kakuyanagi, S.-i. Karimoto, H. Nakano, W. J. Munro, Y. Tokura, M. S. Everitt, K. Nemoto, M. Kasu, N. Mizuochi, and K. Semba, Nature **478**, 221 (2011).

[20]    D. Marcos, M. Wubs, J. M. Taylor, R. Aguado, M. D. Lukin, and A. S. Sorensen, Phys. Rev. Lett. **105**, 210501 (2010).

[21]    J. Koch, T. M. Yu, J. Gambetta, A. A. Houck, D. I. Schuster, J. Majer, A. Blais, M. H. Devoret, S. M. Girvin, and R. J. Schoelkopf, Phys. Rev. A **76**, 042319 (2007).

[22]    F. Jelezko and J. Wrachtrup, Phys. Stat. Sol. (a) **203**, 3207 (2006).

[23]    Y. Nakamura, Y. A. Pashkin, and J. S. Tsai, Nature **398**, 786 (1999).

[24]    M. D. Hutchings, J. B. Hertzberg, Y. Liu, N. T. Bronn, G. A. Keefe, J. M. Chow, and B. L. T. Plourde, Arxiv, 1702.02253.

[25]    R. Meservey and P. M. Tedrow, J. Appl. Phys. **42**, 51 (1971).

[26]    S. Putz, A. Angerer, D. O. Krimer, R. Glattauer, W. J. Munro, S. Rotter, J. Schmiedmayer, and J. Majer, Nature Photon. **11**, 36 (2017).

[27]    C. Greze, Toward a Spin Ensemble Quantum Memory for Superconducting Qubits, Dissertation for the Degree of Doctor of Philosophy, University of Paris VI, France, 2015.

[28]    Y. Kubo, F. R. Ong, P. Bertet, D. Vion, V. Jacques, D. Zheng, A. Dreau, J. F. Roch, A. Auffeves, F. Jelezko, J. Wrachtrup, M. F. Barthe, P. Bergonzo, and D. Esteve, Phys. Rev. Lett. **105**, 140502 (2010).

[29]    S. J. Weber, Quantum Trajectories of a Superconducting Qubit, Dissertation for the Degree of Doctor of Philosophy, University of California, Berkeley, 2014.

[30]    N. Mizuochi, P. Neumann, F. Rempp, J. Beck, V. Jacques, P. Siyushev, K. Nakamura, D. J. Twitchen, H. Watanabe, S. Yamasaki, F. Jelezko, and J. Wrachtrup, Phys. Rev. B **80**, 041201(R) (2009).

[31]     F. Dolde, H. Fedder, M. W. Doherty, T. Nöbauer, F. Rempp, G. Balasubramanian, T. Wolf, F. Reinhard, L. C. L. Hollenberg, F. Jelezko, and J. Wrachtrup, Nature Phys. **7**, 459 (2011).

[32]     C. Grezes, B. Julsgaard, Y. Kubo, M. Stern, T. Umeda, J. Isoya, H. Sumiya, H. Abe, S. Onoda, T. Ohshima, V. Jacques, J. Esteve, D. Vion, D. Esteve, K. Mølmer, and P. Bertet, Phys. Rev. X **4**, 021049 (2014).

[33]     J. R. Maze, A. Dréau, V. Waselowski, H. Duarte, J. F. Roch, and V. Jacques, New J. Phys. **14**, 103041 (2012).

[34]     C. Grezes, B. Julsgaard, Y. Kubo, W. L. Ma, M. Stern, A. Bienfait, K. Nakamura, J. Isoya, S. Onoda, T. Ohshima, V. Jacques, D. Vion, D. Esteve, R. B. Liu, K. Mølmer, and P. Bertet, Phys. Rev. A **92**, 020301 (2015).

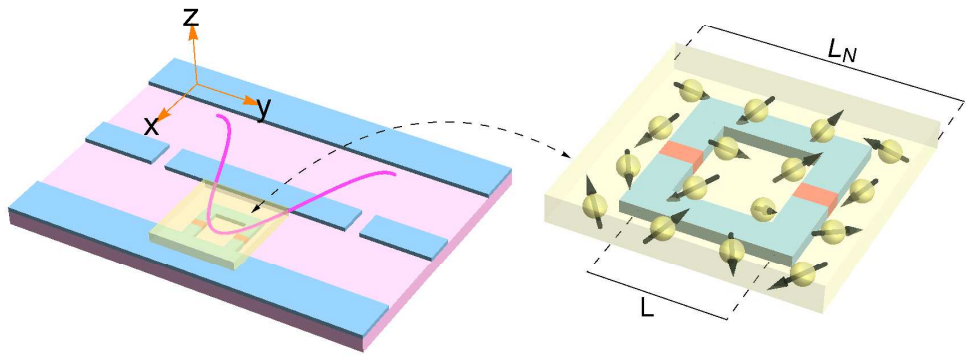

**FIG. 1.** A cavity-transmon-spin-ensemble hybrid system. A transmon is located in the place with a maximum electric field inside a coplanar waveguide (CPW) superconducting cavity. The transmon is covered by a diamond chip. The pink part is the substrate and blue one represents the superconductor film. The purple line is the electric field in the cavity and the red part in zoom area illustrates the insulator barrier of a Josephson junction. The diamond chip is denoted by brown cuboid and the spin of NV center is indicated by black arrows in this figure. The size of NV center is $L_N$ and the distance between two Josephson junctions in the transmon is L. The cross section of the transmon junction is h × h.

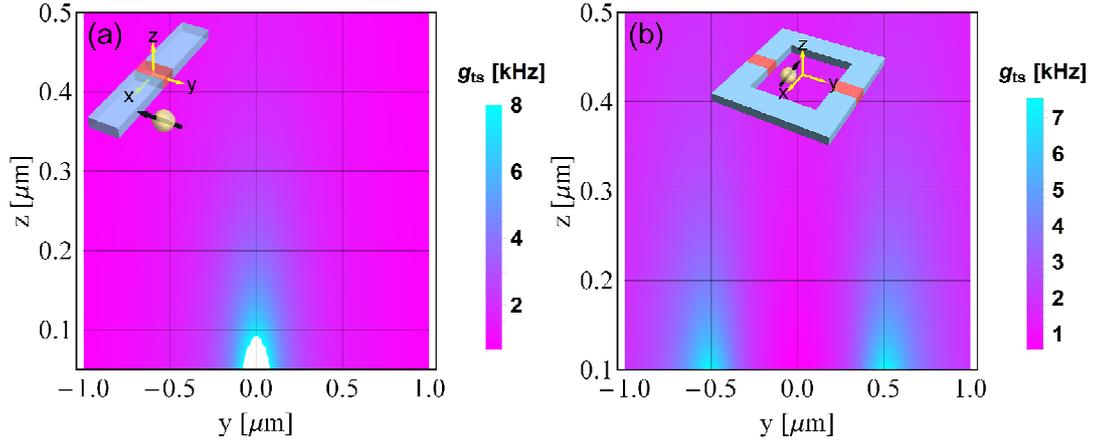

**FIG. 2.** Coupling strength $g_{ts}$ as a function of the location of a single spin with x = 0 for the case (a) a single spin coupling to a single-JJ transmon and (b) a single spin coupling to a double-JJ transmon. Each inset shows the schematic diagram of the coupling scenario. The quantization axis of NV center is assumed to be along the x direction. The critical current of the transmon $I_c$ = 500 nA is used for calculation. The light-white region in (a) represents a coupling strength larger than 8 kHz. The origin of the coordinate in inset (a) is located at the center of the insulator cuboid, so the top surface of transmon is at z = h/2 = 0.05um. For (b), two junctions are indentical and the distance between two junctions is L = 3um. The origin in inset (b) is set at the center of the transmon loop and the top surface of transmon is also at z = h/2 = 0.05um.

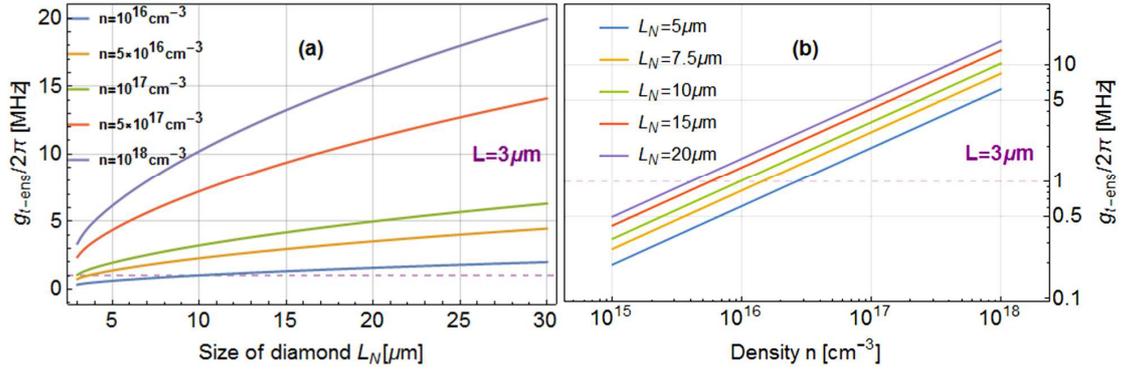

**FIG. 3.** The coupling strength $g_{t-ens}$ between a double-JJ transmon and NV spin ensemble as a function of diamond crystal size $L_N$ with different densities n in (a) and as a function of density n with different diamond crystal dimensions $L_N$ in (b). The diamond crystal has a volume of $L_N^3$. The coupling strength is calculated by summing over all the spins in the diamond cube. The dimension of transmon L (the distance between two junctions) is 3 μm. The purple dashed line represents $g_{t-ens} = 1$ MHz, indicating a strong coupling regime for the transmon-spin ensemble system.

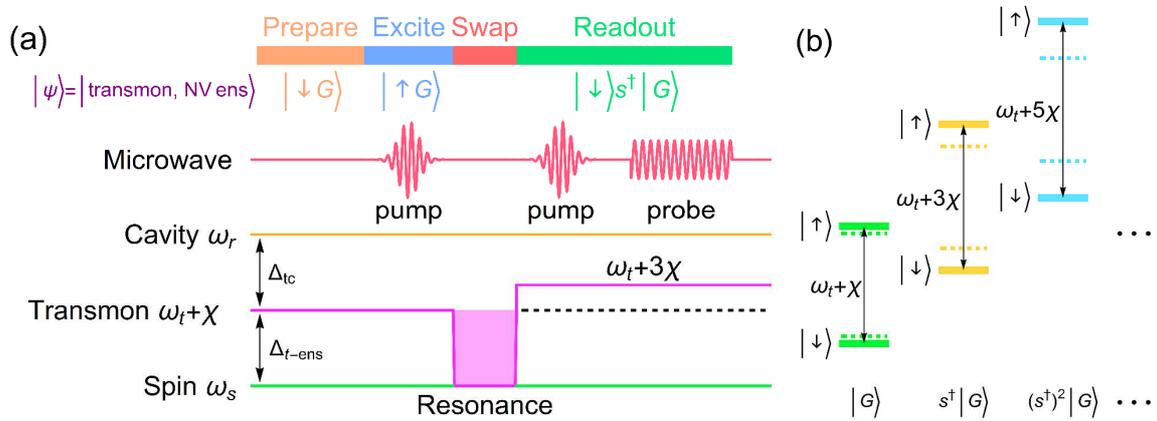

**FIG. 4.** (a) A dispersive readout procedure for the state of NV ensemble. The whole measurement process consists of steps for preparation, excitation, SWAP operation and pump-probe measurement to determine the state of spin ensemble (b) An energy level diagram of the NV ensemble-transmon system. The transition frequency of transmon depends on the state of NV ensemble due to the dispersive interaction between transmon and NV ensemble. The dashed line represents the energy level of transmon-NV-ensemble system without the interaction between them. $|\downarrow\rangle, |\uparrow\rangle$ represents the ground and excited state of transmon and $|G\rangle, s^{\dagger}|G\rangle, (s^{\dagger})^2|G\rangle$ is the ground state and excited states of NV ensemble, respectively.

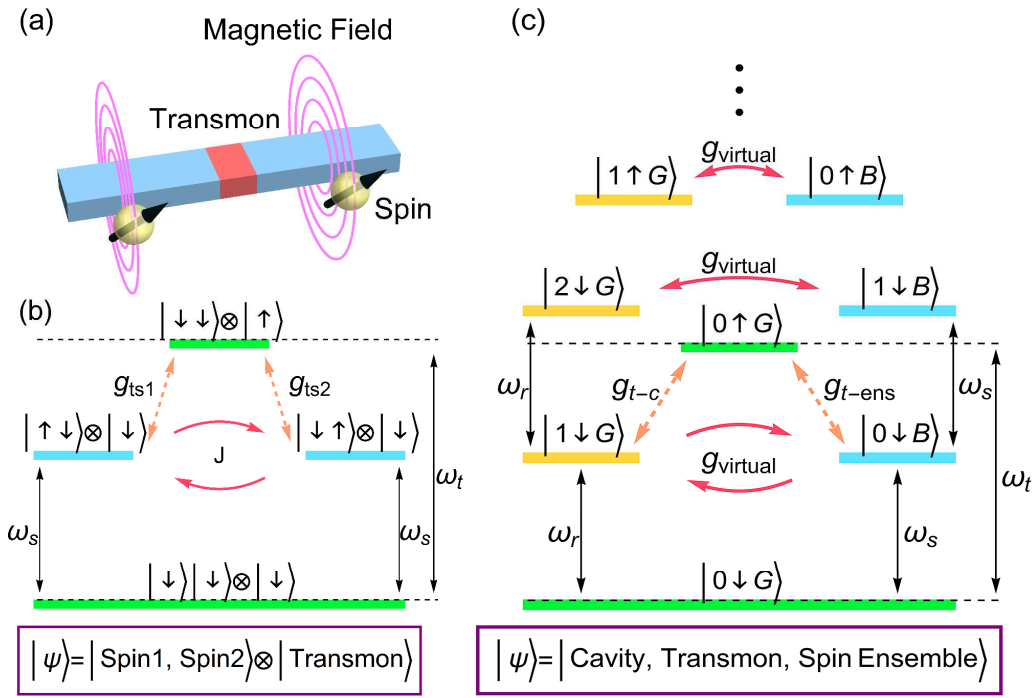

**FIG. 5.** (a) A schematic diagram showing coupling two spins to a single-JJ transmon. The blue part represents the superconductor and red one shows the insulator. Spins are shown as brown spheres with black arrows and purple circles show the magnetic field generated by transmon. (b) Energy level diagram of spin-transmon-spin system. Two energy levels in blue interact with each other via the virtual exchange. Both spins are detuned from the transmon but in resonance with each other to turn on the virtual exchange. (c) Energy level diagram of cavity-transmon-spin ensemble system. The Energy level in yellow interacts with the level in blue to swap an excitation between them. The cavity and spin ensemble are detuned from the transmon to prohibit the real exchange between them.